\documentstyle[aps,preprint,epsfig]{revtex}   

\newcommand{\bea}{\begin{eqnarray}}
\newcommand{\eea}{\end{eqnarray}}
\newcommand{\ba}{\begin{array}}
\newcommand{\ea}{\end{array}}

\newcommand{\vp}{{\vec p}}

\newcommand{\be}{\begin{equation}}
\newcommand{\ee}{\end{equation}}

\newcommand{\bi}{\bibitem}
\newcommand{\wvp}{\vec P}

\newcommand{\ms}{\mathstrut}
\newcommand{\ds}{\displaystyle}
\newcommand{\p}{\partial}
\newcommand{\lsim}{\mathrel{\rlap{\lower5pt\hbox{\hskip0pt$\sim$}}
\raise1pt\hbox{$<$}}}
\tighten
\begin{document}
\preprint{Preprint Number: \parbox[t]{45mm}{ANL-PHY-9373-TH-99}}

\title{Pair creation: back-reactions and damping} 
\author{J.C.R. Bloch$^1$, V.A. Mizerny$^2$, A.V. Prozorkevich$^2$,
C.D. Roberts$^1$,\\ S.M. Schmidt$^1$, S.A. Smolyansky$^2$,
D.V. Vinnik$^2$\vspace*{0.2\baselineskip}}
\address{$^1$Physics Division,
Bldg. 203, Argonne National Laboratory,\\ Argonne, Illinois 60439-4843,
USA\vspace*{0.2\baselineskip}}
\address{$^2$Physics Department, Saratov State University, Saratov, Russian
Federation\vspace*{0.2\baselineskip}}
\date{Pacs Numbers: 05.20.Dd, 25.75.Dw, 05.60.Gg, 12.38.Mh}
\maketitle
\begin{abstract}
We solve the quantum Vlasov equation for fermions and bosons, incorporating
spontaneous pair creation in the presence of back-reactions and collisions.
Pair creation is initiated by an external impulse field and the source term
is non-Markovian.  A simultaneous solution of Maxwell's equation in the
presence of feedback yields an internal current and electric field that
exhibit plasma oscillations with a period $\tau_{pl}$.  Allowing for
collisions, these oscillations are damped on a time-scale, $\tau_r$,
determined by the collision frequency.  Plasma oscillations cannot affect the
early stages of the formation of a quark-gluon plasma unless $\tau_r \gg
\tau_{pl}$ and $\tau_{pl} \sim 1/\Lambda_{\rm QCD} \sim 1\,$fm/c.
\end{abstract}

%
%
\section{Introduction}\label{intro}
The nonperturbative Schwinger mechanism\cite{Sch}, which describes the
spontaneous formation of fermion-antifermion pairs, has been used\cite{agi}
to model the formation of a quark-gluon plasma in heavy ion collisions.  In
this approach\cite{nussinov} nucleon-nucleon collisions lead to the creation
of flux-tubes, in which quark-antiquark pairs are connected by a strong
colour-electric field.  The energy density (string tension) acts like a
strong background field and particle-antiparticle pairs are created via the
Schwinger mechanism\cite{bialas,matsui,knoll,NN}.  These charged particles
polarise the vacuum and are accelerated in the external field.  Their motion
generates a field that in turn modifies the initial background field and, in
the absence of further interactions, that back-reaction induces plasma
oscillations.

The back-reaction phenomenon has become a focus of attention in recent years,
both in general and as it can arise in the pre-equilibrium stage of an heavy
ion collision.  Theoretical approaches as diverse as field
theory\cite{Back,BackA,Rau} and transport equations\cite{KM,GKM} have been
applied.  The link between treatments based on the field equations and the
formulation of a Boltzmann equation was recently investigated\cite{gsi,kme}.
These studies show that the resulting kinetic equation has a non-Markovian
source term.  For weak fields there is no overlap between the time-scales
characterising vacuum tunnelling and the period between pair production
events: $\tau_{qu}$, $\tau_{prod}$, and the Markovian approximation to the
quantum Vlasov equation is valid\cite{kme}.  However, for strong background
fields there is an overlap between these time-scales and this makes the
non-Markovian nature of the source term very
important\cite{basti,basti1,PRD}.  Back-reactions and collisions introduce at
least two more time-scales: the plasma oscillation period, $\tau_{pl}$, and
the collision period, $\tau_{r}$, and their impact is an integral focus of
this article.  Furthermore, in contrast to other recent studies\cite{KM,kme},
we induce particle production by a time-dependent external field.

In Sect.~II we review the main equations and results for particle creation
using a non-Markovian source term.  In Sect.~III we derive the renormalised
Maxwell equation determined by the external and internal fields and, for
special choices of the external field, present numerical results obtained by
solving the coupled system of kinetic and Maxwell equations for bosons and
fermions, with and without a simple collision term.  We summarise our results
in Sect.~IV.

\section{Pair creation with a non-Markovian source term}
 
We consider an external, spatially-homogeneous, time-dependent vector
potential $A_\mu$, in Coulomb gauge: $A_0 =0$, and write
$\vec{A}=(0,0,A(t))$.  The corresponding electric field is
\be
E(t)=-{\dot A}\ms(t) :=-dA(t)/dt\,.
\ee  
The kinetic equation satisfied by the single-particle distribution function:
$f_{\pm}$ (``$+$'' for bosons, ``$-$'' for fermions) is
\begin{equation}
\label{10}
\frac{df_\pm(\vp,t)}{dt}=S_\pm(\vp,t), 
\end{equation}
where the source term is momentum- and time-dependent:
\be\label{source}
S_\pm(\vp,t) = \frac{1}{2}{\cal W}_\pm(t)\int_{-\infty}^t dt'{\cal
W}_\pm(t')F_\pm(\vp,t)  \cos[x(t',t)]\,,
\ee
with $x(t',t) := 2[\Theta(t)-\Theta(t')]$ describing the difference between
the dynamical phases
\be
\label{30}
\Theta(t) = \int^t_{-\infty}dt'\omega(t')\,\,.
\ee
Here the total energy is
\be
\omega(t)=\sqrt{\varepsilon_\perp^2+P_\parallel^2(t)}\,
\ee
where $\varepsilon_\perp=\sqrt{m^2+\vp_\perp^{\,2}}$ is the transverse energy
and we have introduced the kinetic momentum: $\wvp=(p_\perp,P_\parallel(t))$,
with $\vp_\perp = (p_1,p_2)$, $P_\parallel(t)=p_\parallel-eA(t)$.

Equation~(\ref{10}) was recently derived from the underlying quantum field
theory\cite{gsi,kme,basti} and exhibits a number of interesting new features.
For example, in a strong background field its solutions describe an
enhancement in the boson production rate and a suppression of fermion
production.  There are two aspects\cite{basti1,PRD} of Eq.~(\ref{10}) that
generate such differences between the solutions for fermions and bosons: the
different transition coefficients
\be\label{12}
{\cal W}_\pm(t)=\frac{eE(t)P_\parallel(t)}{\omega^2(t)}
\bigg(\frac{\varepsilon_\perp}{P_\parallel(t)}\bigg)^{g_\pm-1}\,,
\ee
where the degeneracy factor is $g_+=1$ for bosons and $g_-=2$ for fermions;
and the statistical factor: $F_\pm(\vp,t) = [1\pm2f_\pm(\vp,t)]$.  

The kinetic equation, Eq.~(\ref{10}), is non-Markovian for two reasons: (i)
the source term on the right-hand-side (r.h.s) requires knowledge of the
entire history of the evolution of the distribution function, from
$t_{-\infty}\rightarrow t$; and (ii), even in the low density limit
($F(t)=1$), the integrand is a nonlocal function of time as is apparent in
the coherent phase oscillation term: $\cos[x(t',t)]$.  The mean field
approaches of Refs.\cite{Back,BackA,Rau} also incorporate non-Markovian
effects in particle production.  However, the merit of a kinetic formulation
lies in the ability to make a simple and direct connection with widely used
approximations.

In the low density limit the source term is independent of the distribution
function \be S^0_\pm(\vp,t)= \frac{1}{2}{\cal W}_\pm(t)\int_{-\infty}^t
dt'{\cal W}_\pm(t') \cos[x(t',t)] \ee
and Eq.~(\ref{10}) becomes
\be\label{100}
\frac{df^0_\pm(\vp,t)}{dt}=S_\pm^0(\vp,t)\,. 
\ee
(The low-density limit can only be self-consistent for weak fields.)  Even in
this case there are differences between the solutions for fermions and bosons
because of the different coefficients ${\cal W}_\pm(t)$, and the equation
remains nonlocal in time.  Equation~(\ref{100}) has the general solution
\be
f^0_\pm(\vp,t)=\int_{-\infty}^t dt'S^0_\pm(\vp,t'),
\ee
which provides an excellent approximation to the solution of the complete
equation when the background field strength is small compared to the
transverse energy.

The ideal Markov limit was found in Ref.\cite{kme}, where a further
asymptotic expansion was employed and a {\bf local} source term for weak
electric fields was derived.  In this case $\tau_{qu} < \tau_{prod}$.
However, for very strong fields a clear separation of these time-scales is
not possible and the kinetic equation must be solved in its non-Markovian
form where memory effects are important\cite{PRD}.

Equation (\ref{10}) is an integro-differential equation.  It can be
re-expressed by introducing
\bea\label{3.1}
v_\pm(\vp,t)&=&\int_{t_0}^t dt'\, {\cal W}_\pm(\vp,t') F_\pm(\vp,t')\,\cos
[x(\vp,t,t')]\,, \\\label{3.11}
z_\pm(\vp,t)&=&\int_{t_0}^t dt'\, {\cal W}_\pm(\vp,t') F_\pm(\vp,t')\,\sin 
[x(\vp,t,t')]\,,
\eea
in which case we have
\bea
\label{f}
\frac{\p f_\pm(\wvp,t)}{\p t}+eE(t)\frac{\p f_\pm(\wvp,t)}
{\p P_\parallel}&=& \case{1}{2} {\cal W}_\pm(\wvp,t') v_\pm(\wvp,t),\\
\label{v}
\,\frac{\p v_\pm(\wvp,t)}{\p t} + eE(t)\,\frac{\p v_\pm(\wvp,t)}
{\p P_\parallel}&=&\ds {\cal W}_\pm(\wvp,t') 
F_\pm(\wvp,t)-2\omega(\wvp) z_\pm(\wvp,t) \,,
\\
\label{z}
\frac{\p z_\pm(\wvp,t)}{\p t}+eE(t)\frac{\p z_\pm(\wvp,t)}{\p
P_\parallel}&=&2\omega(\wvp) v_\pm(\wvp,t) \, ,
\eea
with the initial conditions $f_\pm(t_0)= v_\pm(t_0)=z_\pm(t_0)=0 $, where
$t_0\rightarrow -\infty$.  This coupled system of linear differential
equations is much simpler to solve numerically.

\section{Back-reactions}\label{sec:3}
\subsection{The Maxwell equation}
In recent years the effect of back-reactions in inflationary cosmology and
also in the evolution of a quark-gluon plasma has been studied extensively.
In both cases the particles produced by the strong background field modify
that field: in cosmology it is the time-dependent gravitational field, which
couples via the masses, and in a quark-gluon plasma, it is the chromoelectric
field affected by the partons' colour charge.

In our previous studies we have considered constant\cite{PRD} and
simply-constructed time-dependent\cite{basti} Abelian electric fields but
ignored the effect of back-reactions; i.e., that the particles produced by
the background field are accelerated by that field, generating a current that
opposes and weakens it, and can also lead\cite{Back,BackA} to plasma
oscillations.  The effect of this back-reaction on the induced field is
accounted for by solving Maxwell's equation: $\dot E(t)=-j(t)$.  Herein we
assume that the plasma is initially produced by an external field,
$E_{ex}(t)$, excited by an external current, $j_{ex}(t)$, such as might
represent a heavy ion collision and this is our model-dependent input.  The
total field is the sum of that external field and an internal field,
$E_{in}(t)$, generated by the internal current, $j_{in}(t)$, that
characterises the behaviour of the particles produced.  Hence the total field
and the total current are given by
\bea\label{2.10} 
E(t)&=&E_{in}(t)+E_{ex}(t)\ ,\\
j(t)&=&j_{in}(t)+j_{ex}(t)\ .\label{2.11}
\eea

Continued spontaneous production of charged particle pairs creates a
polarisation current, $j_{pol}(t)$, that depends on the particle production
rate, $S(\vp,t)$.  Meanwhile the motion of the existing particles in the
plasma generates a conduction current, $j_{cond}(t)$, that depends on their
momentum distribution, $f(\vp,t)$.  The internal current is the sum of these
two contributions
\be\label{1.14}
{\dot E}_{in}(t)=-j_{in}=-j_{cond}(t)-j_{pol}(t)\,.
\ee
At mean field level the currents can be obtained directly from the constraint
of local energy density conservation: $\dot \epsilon = 0$, where
\begin{equation}
\epsilon(t) = \frac{1}{2}E^2(t)
+ 2\int \frac{d^3 \vp}{(2\pi)^3}\omega(\vp,t)f(\vp,t)\,.
\end{equation}
For bosons this constraint yields
\be
\label{eEdiv}
{\dot E}(t) = -2e\int\frac{d^3
\vp}{(2\pi)^3}\frac{p_\parallel-eA(t)}
{\omega(\vp,t)}\bigg[f(\vp,t)+\frac{\omega(\vp,t)}{{\dot
\omega}(\vp,t)}\frac{df(\vp,t)}{dt}\bigg]\,, 
\ee
and we can identify the conduction current
\be\label{jin}
j_{cond}(t)=2e\int\frac{d^3
\vp}{(2\pi)^3}\frac{p_\parallel-eA(t)}{\omega(\vp,t) } 
f(\vp,t)\ ,\label{jcond}
\ee
and, using Eq.~(\ref{10}), the polarisation current
\be
j_{pol}(t)=\frac{2}{E(t)}\int\frac{d^3 p}{(2\pi)^3}\,\omega(\vp,t)S(\vp,t)
\label{jpol}\,. 
\ee
Thus, using Eqs.~(\ref{3.1}) and (\ref{3.11}), Maxwell's equation is
\bea\label{dotE}
\dot E_{in}(t) = -\ddot A_{in}(t)=
-2e\int \frac{d^3\wvp}{(2\pi)^3}\frac{P_\parallel(t)}{\omega(\wvp)}
\bigg[f(\wvp,t)+\frac{1}{2}v(\wvp,t)\bigg]\,.
\eea

It is important to observe that this form for the internal field has been
employed extensively in the study of back-reactions.  However, our
contribution is to employ it in conjunction with a time-dependent {\bf
external field}, which allows for the exploration of a richer variety of
phenomena.
 
\subsection{Renormalisation}
The boson and fermion currents are 
\be\label{jinn}
j_{in}(t)=
eg_\pm\int\frac{d^3\wvp}{(2\pi)^3}
\frac{P_\parallel(t)}{\omega(\wvp)}\bigg[f(\wvp,t)
+\frac{v(\wvp)}{2}\bigg(\frac{\epsilon_\perp}
{P_\parallel(t)}\bigg)^{g_\pm-1}\bigg]\,,
\ee
where the integrand depends on the solution of the kinetic equation,
Eqs.~(\ref{f})-(\ref{z}), which must be such as to ensure the integral is
finite.  Simple power counting indicates that admissible solutions must
satisfy
\begin{equation}
v(\wvp,t)\,,\;f(\vec{P},t) \stackrel{|\vec{P}| \to \infty}{\lsim}
\frac{1}{|\vec{P}|^4} \,.
\end{equation}

To fully characterise the asymptotic behaviour we employ a separable
Ansatz
\begin{equation}
\label{sums}
f(\wvp,t)= \sum^\infty_{k=0}\frac{f_k(t)}{|\vec{P}|^k}\,, \;
v(\wvp,t)= \sum^\infty_{k=0}\frac{v_k(t)}{|\vec{P}|^k}\,, \;
z(\wvp,t)= \sum^\infty_{k=0}\frac{z_k(t)}{|\vec{P}|^k}\,.
\end{equation}
Substituting these in Eqs.~(\ref{f})-(\ref{z}) and comparing coefficients,
using $P_\| \approx \omega(\vec{P}) \approx \epsilon_\perp$, which are valid
at large $|\vec{P}|$, we find the leading terms
\begin{equation}
\label{2223}
f_4  =
\case{1}{16} e^2 E^2(t) \,,\;
v_3  =  \case{1}{4} e{\dot E}(t) \,,\;
z_2 =  \case{1}{2} eE(t) \,,
\end{equation}
with all the lower-order coefficients being zero.  Substituting these results
in Eq.~(\ref{jin}) it is clear that the conduction current is convergent.
However, there is a logarithmic divergence in the polarisation current,
Eqs.~(\ref{eEdiv}) and (\ref{jpol}), which is apparent in Eq.~(\ref{jinn}),
but that is just the usual short-distance divergence associated with charge
renormalisation.  We regularise the polarisation current by writing $v = (v -
v_3 P_\|/\omega^4) + v_3P_\|/\omega^4$, so that
\begin{eqnarray}
\label{Ereg}
\lefteqn{{\dot E}^\pm(t) = -j_{ex}(t)}\\
&& \nonumber
-g_\pm e\int\frac{d^3\wvp}{(2\pi)^3}\frac{P_\parallel(t)}{\omega(\wvp)}
\left[f(\wvp,t)+
\case{1}{2}
\left\{
v(\wvp,t)-\frac{e \dot E(t)\,P_\parallel(t) }{4\,\omega^4(\vec{P})}
\right\}
\bigg(\frac{\epsilon_\perp}{P_\parallel(t)}\bigg)^{g_\pm-1}
\right]
- e^2{\dot E}^\pm(t) I^\pm(\Lambda)\,,
\end{eqnarray}
where
\begin{equation}
\label{regB}
I^\pm(\Lambda) =  \frac{g_\pm}{4}\,\int\frac{d^3\wvp}{(2\pi)^3}
\frac{P^2_\parallel(t) }{\omega^5(\vec{P})}
\left(\frac{\epsilon_\perp}{P_\parallel(t)}\right)^{g_\pm-1}
\stackrel{\Lambda\to\infty}{=} \frac{g_\pm}{8\pi^2} 
\ln\left[\Lambda^2/m^2\right]\,,
\end{equation}
with $\Lambda$ a cutoff on $|\vec{P}|$, which effects a regularisation
equivalent\cite{BackA} to Pauli-Villars.  Introducing the renormalised
charge, fields and current: 
\begin{equation}
e^2_R  = Z\, e^2 \,, \;
{\cal E}^\pm(t)  =  E^\pm(t)/\sqrt{Z}\,,\;
{\cal A}^\pm(t)  =  A^\pm(t)/\sqrt{Z}\,,\;
{\cal J}_{ex}(t) = \sqrt{Z}\,j_{ex}(t) \,,
\end{equation}
with $Z=1/(1+e^2\,I^\pm(\Lambda))$, and noting that $ eE^\pm(t) =
e_R\,{\cal E}^\pm(t)$ and $eA^\pm(t)=e_R\,{\cal A}^\pm(t)$, Eq.~(\ref{Ereg})
becomes
\begin{eqnarray}
\lefteqn{-{\ddot {\cal A}}^\pm(t)  = 
{\dot {\cal E}}^\pm(t)= - {\cal J}_{ex}(t)}\\
\nonumber && 
-g_\pm e_R \int\frac{d^3\wvp}{(2\pi)^3}\frac{P_\parallel(t)}{\omega(\wvp)}
\left[f_\pm(\wvp,t)
+ \case{1}{2}\left\{
v_\pm(\wvp,t)-\frac{e_R\, \dot {\cal E}^\pm(t)\,P_\parallel(t) }
        {4\,\omega^4(\vec{P})}
\right\}
\bigg(\frac{\epsilon_\perp}{P_\parallel(t)}\bigg)^{g_\pm-1}\right]\,.
\end{eqnarray}
This defines a properly renormalised equation for the fields.  Our procedure
is technically different from that employed elsewhere\cite{BackA,cooper} but
yields an equivalent result.  Subsequently all fields and charges are to be
understood as renormalised.
  
\subsection{Numerical results}
Equations~(\ref{f})-(\ref{z}) together with Maxwell's equation,
Eq.~(\ref{Ereg}), form a coupled system of differential equations.  To solve
it we first evaluate the internal current from Eq.~(\ref{Ereg}) at the
primary time-slice using the initial conditions for the distribution
function.  That, via Eqs.~(\ref{2.10}) and (\ref{2.11}), provides an electric
field, which we use to calculate the momentum distribution from
Eqs.~(\ref{f})-(\ref{z}).  This procedure is repeated as we advance over our
time-grid.  We use a momentum grid with $200$ transverse- and $400$
longitudinal-points, a time-step $dt=0.005$, and $\Lambda = 50$ in
Eq.~(\ref{regB}).  All dimensioned quantities are expressed in units of $m$,
the parton mass.

Spontaneous particle creation occurs in the presence of a strong field under
whose influence the vacuum becomes unstable and decays.  Herein we induce
this by a time-dependent external field and compare three different field
configurations.  The fields vanish at $t \to-\infty$, and at $t=t_0$ the
magnitude of the field increases and eventually leads to particle creation.
Configuration (i): For comparison with Refs.\cite{Back,BackA,kme}, we solve
the set of equations as an initial value problem without an external field,
using an initial value of the electric field that is large enough to cause
pair production.
Configuration (ii):  We employ
\be
 \label{step} 
A_{ex}=-A_{0}b^2 \,[t/b+\ln(2\cosh(t/b))],
\hspace{2cm} E_{ex}(t)=A_0 b \,[\tanh(t/b)+1]\,,
\ee
which is an electric field that ``switches-on'' at $t\sim -b$ and evolves to
a constant value, $2 A_0 b$, in an interval $t\sim 2/b$.
Configuration (iii): Is an impulse field configuration:
\be\label{impulse}
A_{ex}(t)=A_0[\tanh(t/b)+1],\hspace{2cm}
E_{ex}(t)=-A_0[b\cosh^{2}(t/b)]^{-1},  \ee 
which is an electric field that ``switches-on'' at $t\sim -2 b$ and off at
$t\sim 2 b$, with a maximum magnitude of $A_0/b$ at $t=0$.  Once this field
has vanished only the induced internal field remains to create particles and
affect their motion.

\begin{figure}[bt]
\epsfig{figure=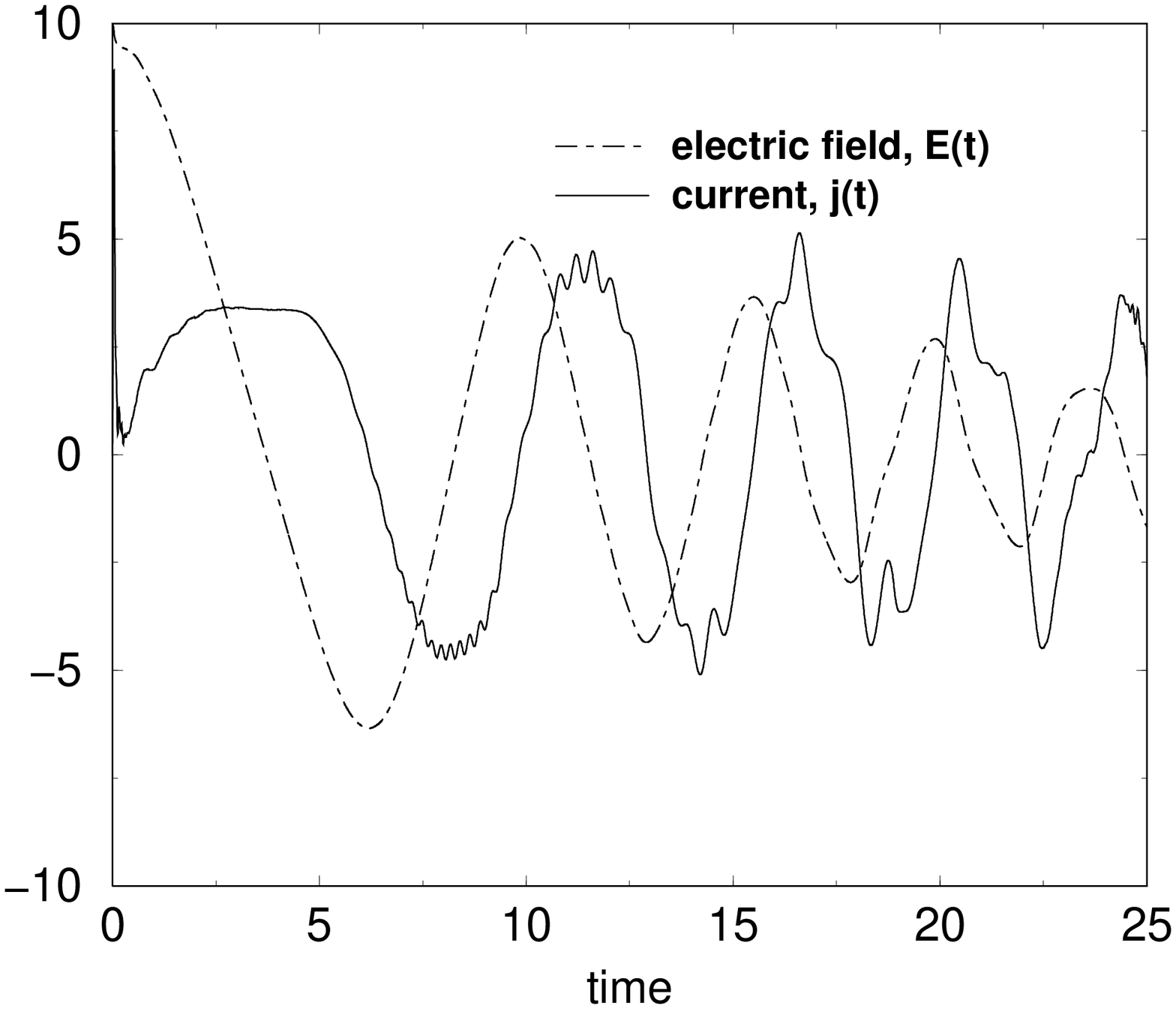,width=15cm,height=15cm,angle=-90}
\caption{Time evolution for bosons of the total electric field and total
current with initial values $E = 10$ and $e^2 = 4$.  (Here, as in the text,
all dimensioned quantities are given in units of the mass-scale
$m$.)\label{fig1}}
\end{figure}

\begin{figure}[bt]
\epsfig{figure=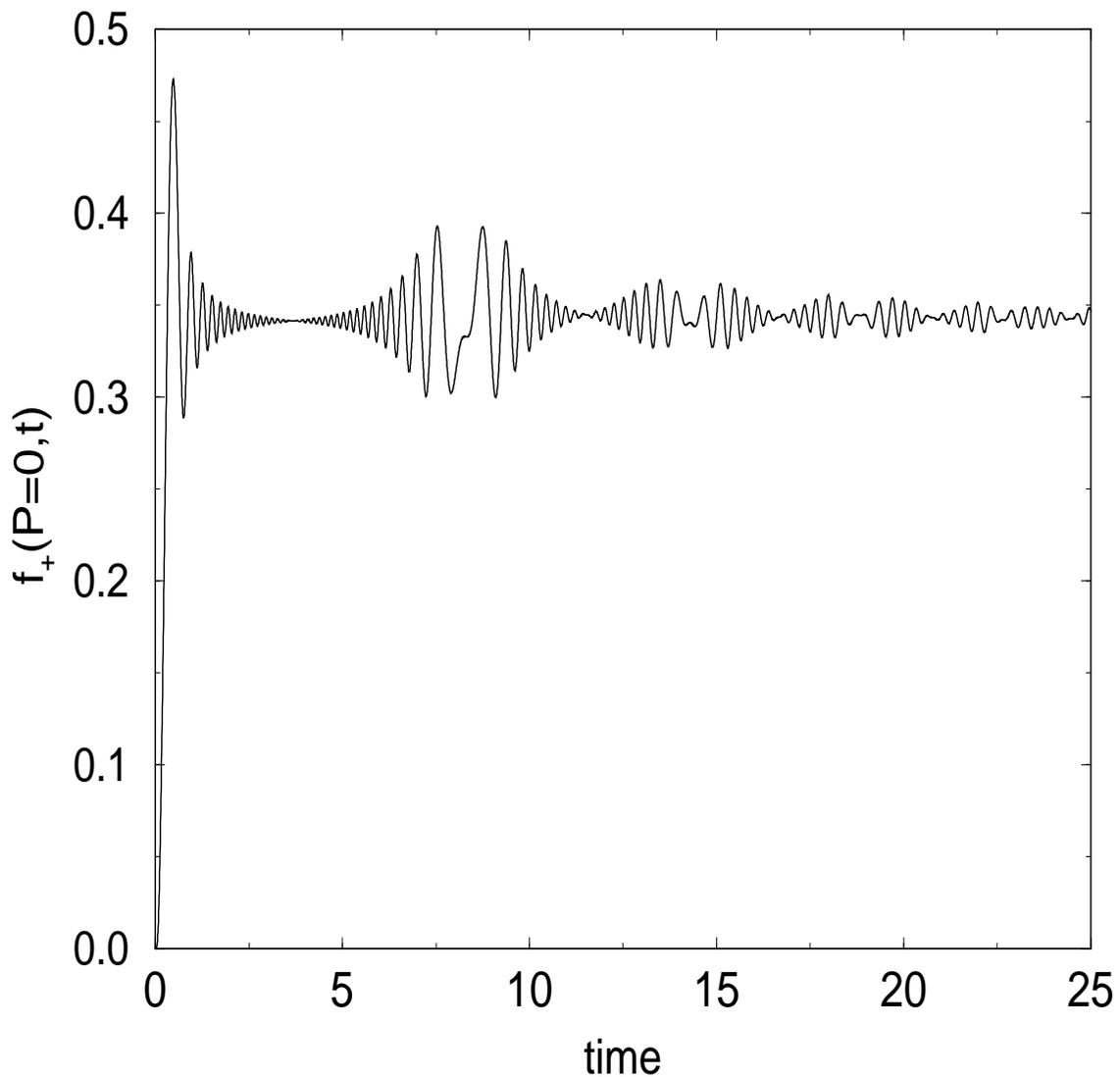,width=15cm,height=15cm,angle=-90}
\caption{Time evolution of the distribution function $f(\bar 0,t)$ with
initial values $E = 10$ and $e^2 = 4$ for bosons.  The rapid fluctuations
occur on the vacuum tunnelling time-scale, $\tau_{qu}$, and coincide with the
{\it Zitterbewegung} identified in the current.  In the absence of
back-reactions these recurrent fluctuation packets are absent\protect\cite{PRD}.
\label{fig3}}
\end{figure}

For configuration (i) we fix an initial value of $E(t=0)=10$, in units of
$m$, with $e^2=4$, and for bosons obtain the electric field and current
depicted in Fig. \ref{fig1}, where plasma oscillations are evident.  The
frequency of these oscillations increases with the magnitude of the field.
The current exhibits a plateau for small $t$, when the particles reach their
maximum velocity.  This current opposes the field, and leads to a suppression
of particle production and a deceleration of the existing particles.  The
effect of this is to overwhelm the field and change its sign with a
consequent change in the direction of the particles' collective motion.  The
process repeats itself, yielding the subsequent oscillations that persist in
the absence of additional interactions, such as collisions or radiation.  The
structure visible at the peaks and troughs of the current is {\bf not} a
numerical artefact.  It is related to the field-strength/mass ratio, being
more pronounced for large values, and occurs on a time-scale $\sim
\tau_{qu}$, the vacuum tunnelling time, and hence can be characterised as
{\it Zitterbewegung}.  It disappears if an ideal-Markovian approximation to
the source term is used because that cannot follow oscillations on such small
time-scales\cite{kme}.  The $t$-dependence of the $\vec{p}=0$ distribution
function is depicted in Fig.~\ref{fig3}, where the beat-like pattern is the
result of back-reactions and the rapid fluctuations coincide with the {\it
Zitterbewegung} identified in the current.

\begin{figure}[h]
\psfig{figure=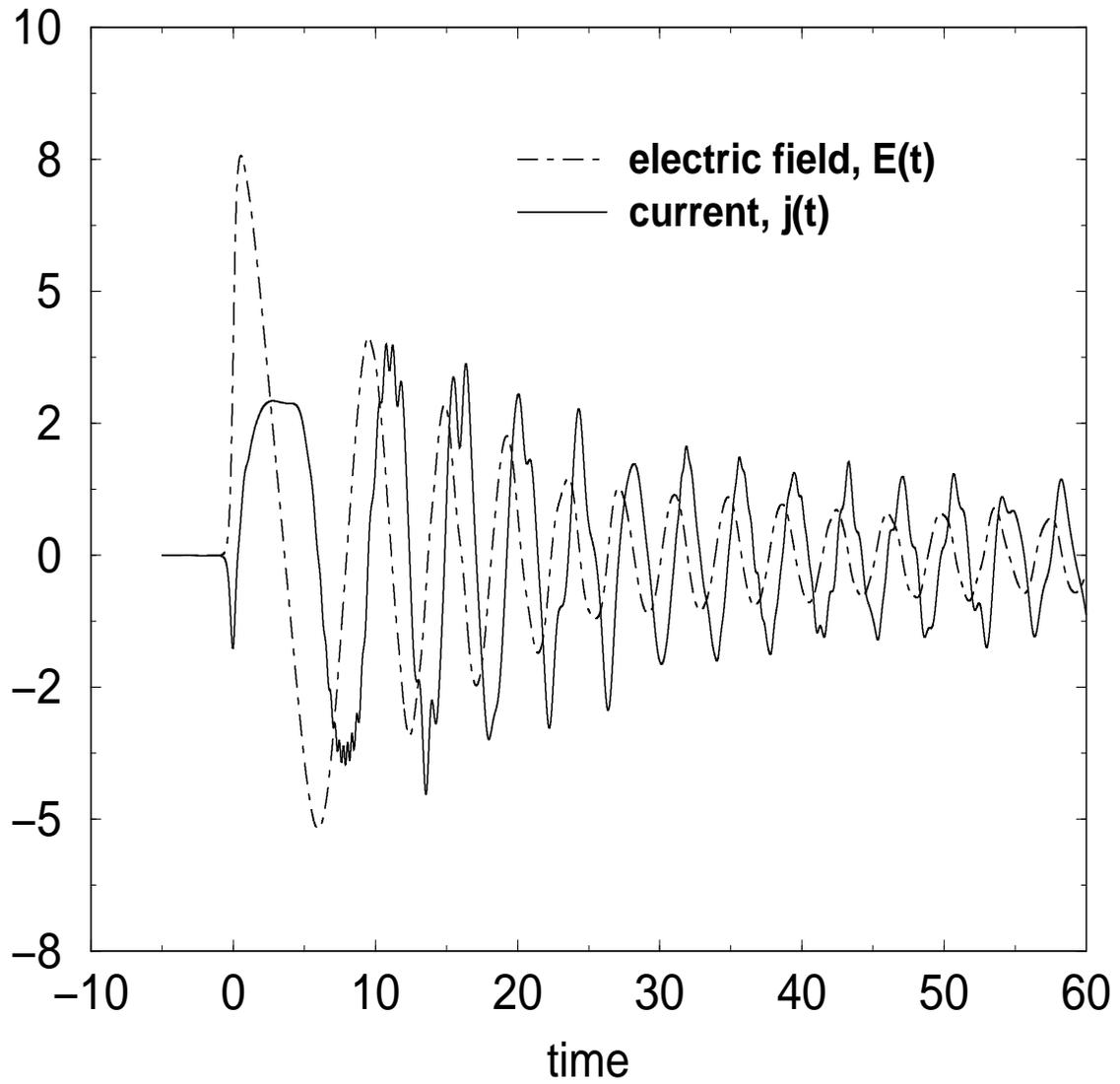,width=15cm,height=15cm,angle=-90}
\caption{Time evolution for bosons of the electric field and the current for
the step function external field, Eq.~(\protect\ref{step}), with $A_0=14.0$,
$b=0.25$ and the coupling $e^2=6$. \label{fig6} }
\end{figure}

\begin{figure}[h]
 \psfig{figure=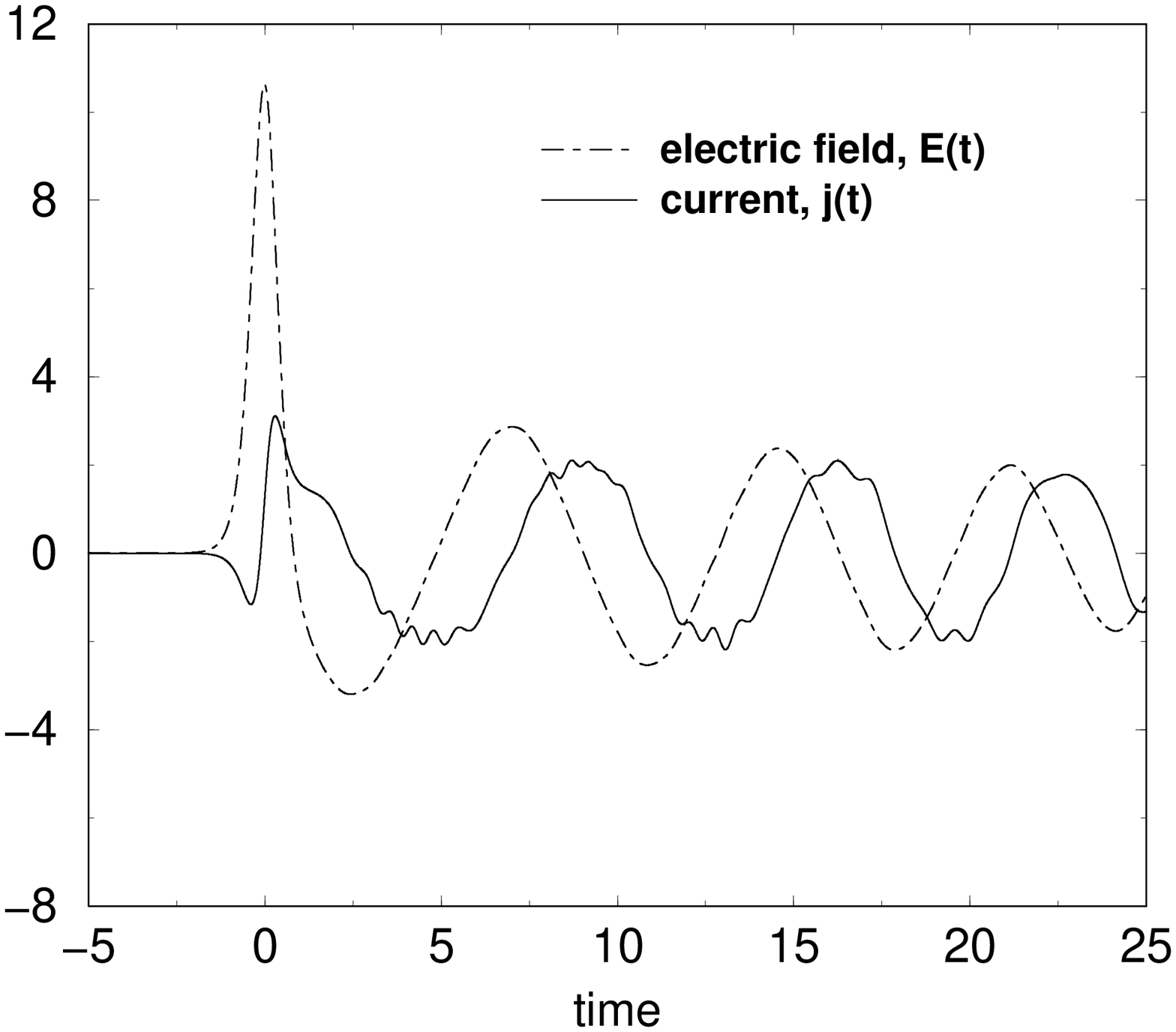,width=15cm,height=15cm,angle=-90}
\caption{Time evolution for bosons of the electric field and current for an
impulse external field, Eq.~(\protect\ref{impulse}), with $A_0=10.0$, $b=0.5$
and the coupling $e^2=5$.\label{fig8} }
\end{figure}

Configurations (ii) and (iii) are alike in that the field ``switches-on'' at
a given time.  However, for (ii) the external field remains constant as $t$
increases whereas in (iii) it ``switches-off'' after $t\sim 2/b$.  The
electric field and current obtained for bosons in these cases are depicted in
Figs.~\ref{fig6} and \ref{fig8}: plasma oscillations are again evident.  We
plot the {\bf total} electric field and thus it is evident in Fig.~\ref{fig6}
that the internal electric field evolves to completely compensate for the
persistent external field, which alone would appear as a straight-line at
$E(t)=7$.  Unsurprisingly, as we see from Fig.~\ref{fig8}, a stable state is
reached more quickly in the absence of a persistent electric field.  Outside
the temporal domain on which the vector potential acts, the initial value and
impulse solutions are equivalent.

\begin{figure}[h]
 \psfig{figure=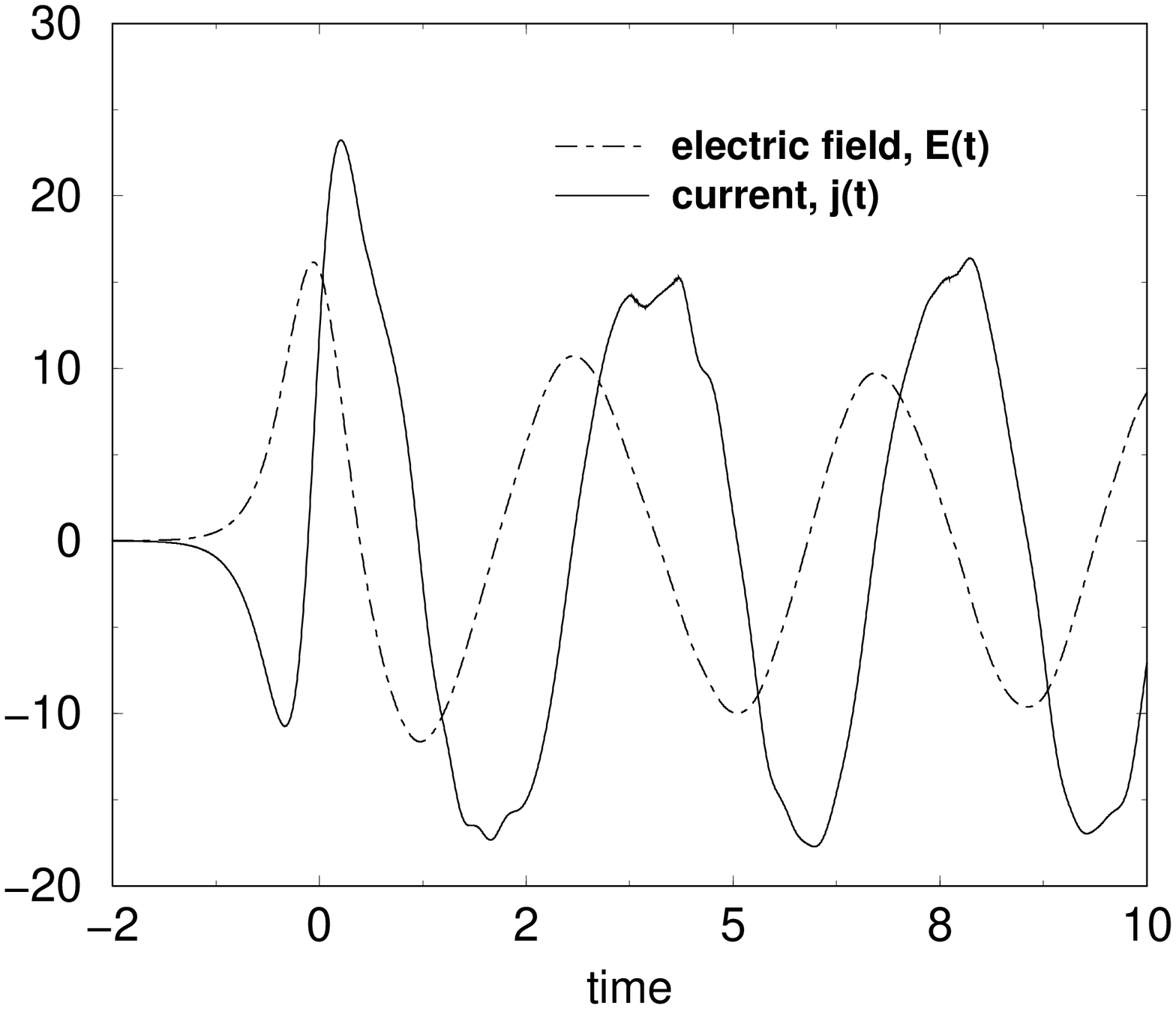,width=15cm,height=15cm,angle=-90}
\caption{Time evolution for fermions of the electric field and current for an
impulse external field, Eq.~(\protect\ref{impulse}), with $A_0=10.0$, $b=0.5$
and the coupling $e^2=4$.\label{fig10} }
\end{figure}

We illustrate the results for fermions in Fig.~\ref{fig10} using the impulse
configuration.  The amplitude and frequency of the plasma oscillations are
significantly larger than for bosons in a configuration of equal strength.
Further, the stable state is reached more quickly because Pauli blocking
inhibits particle production; i.e., no particles can be produced once all
available momentum states are occupied.  Pauli blocking also guarantees
$f_-(\vec{p},t)<1$, for all $t$.

We have also calculated the $P_\|$- and $p_\perp$-dependence of $f$ for both
bosons and fermions.  We find $f_+(t=0)=0$; i.e., bosons cannot be produced
with zero kinetic momentum, an effect readily anticipated from
Eq.~(\ref{12}).  For small $t$, $f_\pm(\vec{p},t)$ is a slowly varying
function of $\vec{p}$ on its domain of support.  However, with increasing
$t$, $f_\pm(\vec{p},t)$ develops large-magnitude fluctuations without
increasing that domain.  The momentum-space position of the midpoint of the
domain of support oscillates with a $t$-dependence given by the kinetic
momentum: $P_\parallel = p_\parallel - eA(t)$.

One additional observation is important here.  The magnitude, $\propto A_0$,
of the electric fields we have considered is large and hence the time between
pair production events, $\tau_{prod}$, is small, being inversely proportional
to the time-average of the source term, $S$.  The period of the plasma
oscillations, $\tau_{pl}$, also decreases with increasing $A_0$ but
nevertheless we always have $\tau_{prod} \ll \tau_{pl}$.  Thus, in contrast
to the effect it has on the production process\cite{PRD}, the temporal
nonlocality of the non-Markovian source term is unimportant to the collective
plasma oscillation.

\subsection{Collisions}
In the previous subsection we ignored the effect of collisions when treating
the spontaneous production of charged particles and subsequent evolution of
the plasma.  Now we consider the effect of a simple collision
term\cite{degrot}
\begin{equation}
\label{collterm}
C_\pm(\vp,t) =\frac{f_\pm^{eq}(\vp,t) - f_\pm(\vp,t)}{\tau_r}\,,
\end{equation}
where $\tau_r$ is the ``relaxation time'' and $f_\pm^{eq}$ are the thermal
equilibrium distribution functions for bosons and fermions:
\be \label{feq} f_\pm^{eq}(\vp,t) =
\frac{1}{\exp[\omega(\vp,t)/T(t)]\mp 1}\,. \ee 
Here $T(t)$ is the ``instantaneous temperature'', which is a model-dependent
concept\cite{matsui,bahl,eis}, and since our results are not particularly
sensitive to details of its form we employ a simple parametrisation
\begin{equation}
T(t)= T_{eq} + (T_m-T_{eq}) \,{\rm e}^{-t^2/t_0^2}\,,
\end{equation}
with an equilibrium temperature $T_{eq}= 1.0$, a maximum temperature
$T_m=2.0$, and a profile-width $t_0^2=10 \sim \tau_{pl}$.  The collision term
is added to the r.h.s.\ of Eq.(\ref{10}), which becomes
\begin{equation}
\label{coll}
\frac{d f_\pm(\vp,t)}{ dt} =S_\pm(\vp,t) + C_\pm(\vp,t)\,.
\end{equation}
This ``relaxation time'' approximation assumes that the system evolves
rapidly towards thermal equilibrium after the particles are produced.  It has
been used before, both in the absence\cite{bialas} of back-reactions and
including them\cite{GKM,bahl,eis}, but with source terms that neglect
fluctuations on short time-scales.

We note that in the low density limit: $f(\vp,t) \ll 1$, one can neglect the
distribution function in the source term, Eq.~(\ref{source}), and
Eq.~(\ref{coll}) has the simple solution
\begin{equation}
f^0_\pm(\vp,t)=
\int_{-\infty}^{\,t}\,dt'\exp\bigg[\frac{t^\prime-t}{\tau_r}\bigg]\bigg(
S_\pm^0(\vp,t')+\frac{f_\pm^{eq}(\vec{p},t^\prime)}{\tau_r}\bigg)\,.
\end{equation}

\begin{figure}[h]
\psfig{figure=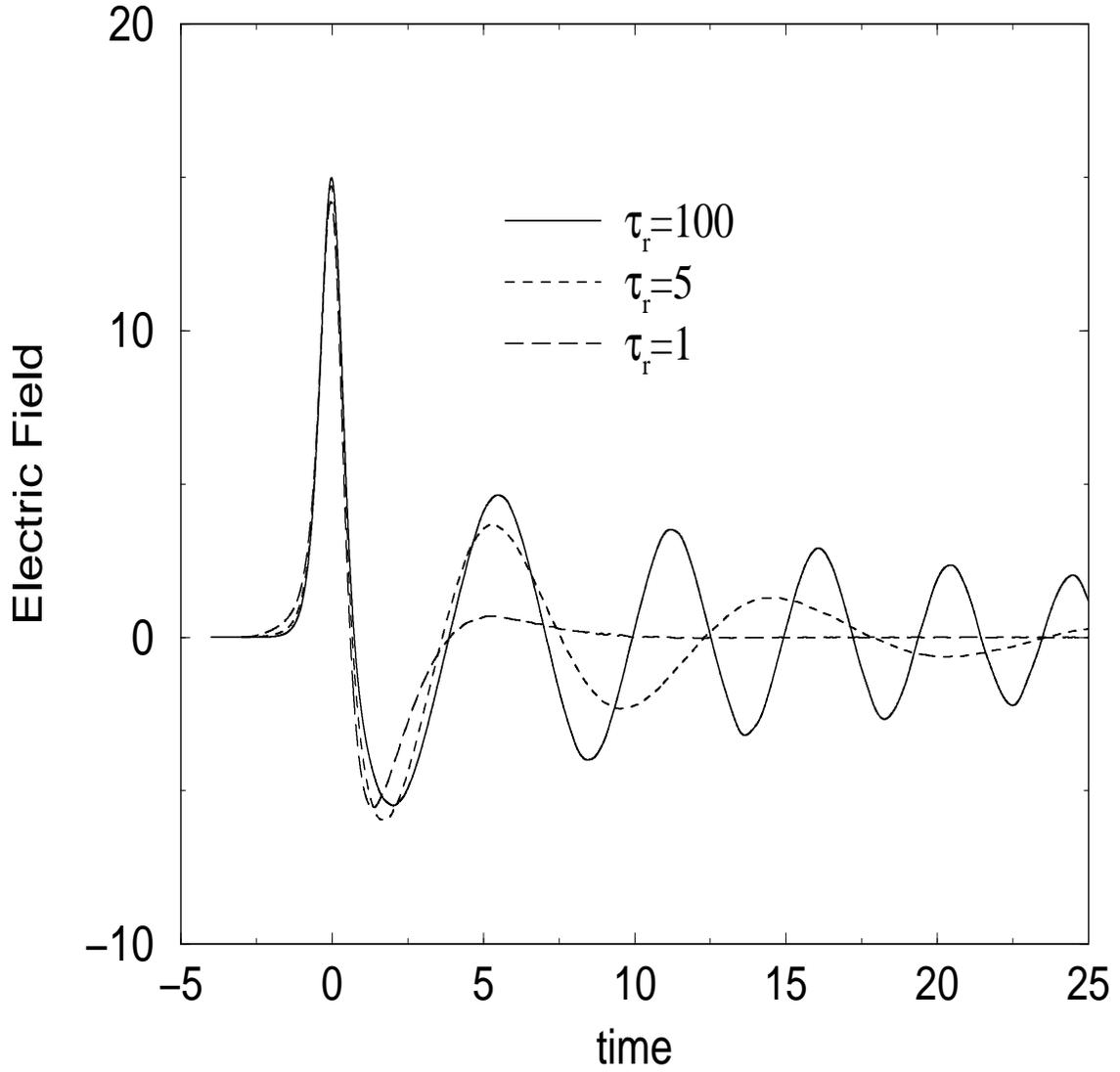,width=15cm,height=15cm,angle=-90}
\caption{Time evolution for bosons of the electric field obtained using
different relaxation times in the collision term of
Eq.~(\protect\ref{collterm}), and with the impulse external field,
Eq.~(\protect\ref{impulse}), where $A_0=7.0$, $b=0.5$ and the coupling
$e^2=4$.}
\label{fig12}
\end{figure}

In our numerical studies we treat $\tau_r$ as a parameter and study the
effect of $C$ on the plasma oscillations.  Our results for bosons using this
crude approximation are depicted in Fig.~\ref{fig12}.  For $\tau_r \gg
\tau_{pl}$ the oscillations are unaffected, as anticipated if $1/\tau_r$ is
interpreted as a collision frequency.  For $\tau_r \sim \tau_{pl}$ the
collision term has a significant impact, with both the amplitude and
frequency of the plasma oscillations being damped.  There is a $\tau_r$ below
which no oscillations arise and the systems evolves quickly and directly to
thermal equilibrium.

\section{Summary and Conclusion}
We have studied spontaneous particle creation in the presence of
back-reactions and collisions, both of which dramatically affect the solution
of the kinetic equation.  The back-reactions lead to plasma oscillations that
are damped by the thermalising collisions if the collision frequency is
comparable to the plasma frequency.  In electric fields where the period of
the plasma oscillations is large compared to the time-scales characterising
particle production, the non-Markovian features of the source term play
little role in the back-reaction process.

Plasma oscillations are a necessary feature of all studies such as ours but
are they relevant to the creation of a quark-gluon plasma?  If we set the
scale in our calculations by assuming that fermions are created in the
impulse configuration with $\langle \epsilon_\perp\rangle \sim \Lambda_{\rm
QCD} \sim 0.5 \sqrt\sigma$, the QCD string tension, then $A_0=10$ with
$e^2=5$ corresponds to an initial field strength $e E \sim 15\,\sigma$ and
energy density $\case{1}{2} E^2 \sim 20\,\sigma^2$.  These are very large
values but even so the plasma oscillation period is still large:
$\tau_{pl}=5\,$fm/c, and collisions can only act to increase that.  We
therefore expect that a quark-gluon plasma will have formed and decayed well
before plasma oscillations can arise.  On these short time-scales
non-Markovian effects will be important.  

Our estimate shows back-reactions to be unimportant on small time-scales but
that is not true of collisions.  However, it is clear that in QCD
applications they must be described by something more sophisticated than the
``relaxation-time'' approximation.  

Finally, with $1/\Lambda_{\rm QCD}$ setting the natural scale, the finite
interaction volume will clearly be important and hence the assumption of a
spatially homogeneous background field must also be improved before
calculations such as ours are relevant to a quark-gluon plasma.
Ref.~\cite{mel} is a step in that direction.

\section*{Acknowledgments}
V.A.M and D.V.V. are grateful for the support and hospitality of the Rostock
University where part of this work was conducted.  This work was supported by
the US Department of Energy, Nuclear Physics Division, under contract
no. W-31-109-ENG-38; the US National Science Foundation under grant
no. INT-9603385, the State Committee of Russian Federation for Higher
Education under grant N 29.15.15; BMBF under the program of
scientific-technological collaboration (WTZ project RUS-656-96); the
Hochschulsonderprogramm (HSP III) under the project No. 0037-6003; and
benefited from the resources of the National Energy Research Scientific
Computing Center. S.M.S. is a F.-Lynen Fellow of the A.v. Humboldt
foundation.



\newpage

\newpage

\newpage

\newpage

\newpage

\end{document}